\documentstyle[psfig]{mn}

\newif\ifAMStwofonts 
 
\ifoldfss 
  \ifCUPmtlplainloaded \else 
    \NewTextAlphabet{textbfit} {cmbxti10} {} 
    \NewTextAlphabet{textbfss} {cmssbx10} {} 
    \NewMathAlphabet{mathbfit} {cmbxti10} {} 
    \NewMathAlphabet{mathbfss} {cmssbx10} {} 
  \fi 
  \ifAMStwofonts 
    \ifCUPmtlplainloaded \else 
      \NewSymbolFont{upmath} {eurm10} 
      \NewSymbolFont{AMSa} {msam10} 
      \NewMathSymbol{\upi}     {0}{upmath}{19} 
      \NewMathSymbol{\umu}     {0}{upmath}{16} 
      \NewMathSymbol{\upartial}{0}{upmath}{40} 
      \NewMathSymbol{\leqslant}{3}{AMSa}{36} 
      \NewMathSymbol{\geqslant}{3}{AMSa}{3E}

       \let\le=\leqslant 
       \let\ge=\geqslant 
    \fi 
  \fi 
\fi 
 
\ifnfssone 
  \newmathalphabet{\mathit} 
  \addtoversion{normal}{\mathit}{cmr}{m}{it} 
  \addtoversion{bold}{\mathit}{cmr}{bx}{it} 
  \newmathalphabet{\mathbfit} 
  \addtoversion{normal}{\mathbfit}{cmr}{bx}{it} 
  \addtoversion{bold}{\mathbfit}{cmr}{bx}{it} 
  \newmathalphabet{\mathbfss} 
  \addtoversion{normal}{\mathbfss}{cmss}{bx}{n} 
  \addtoversion{bold}{\mathbfss}{cmss}{bx}{n} 
  \ifAMStwofonts 
    \ifCUPmtlplainloaded \else 
      \UseAMStwoboldmath 
      \makeatletter 
      \new@mathgroup\upmath@group 
      \define@mathgroup\mv@normal\upmath@group{eur}{m}{n} 
      \define@mathgroup\mv@bold\upmath@group{eur}{b}{n} 
      \edef\UPM{\hexnumber\upmath@group} 
      \new@mathgroup\amsa@group 
      \define@mathgroup\mv@normal\amsa@group{msa}{m}{n} 
      \define@mathgroup\mv@bold\amsa@group{msa}{m}{n} 
      \edef\AMSa{\hexnumber\amsa@group} 
      \makeatother 
      \mathchardef\upi="0\UPM19 
      \mathchardef\umu="0\UPM16 
      \mathchardef\upartial="0\UPM40 
      \mathchardef\leqslant="3\AMSa36 
      \mathchardef\geqslant="3\AMSa3E 

       \let\le=\leqslant 
       \let\ge=\geqslant 
    \fi 
  \fi 
\fi 
 
\ifnfsstwo 
  \DeclareMathAlphabet{\mathbfit}{OT1}{cmr}{bx}{it} 
  \SetMathAlphabet\mathbfit{bold}{OT1}{cmr}{bx}{it} 
  \DeclareMathAlphabet{\mathbfss}{OT1}{cmss}{bx}{n} 
  \SetMathAlphabet\mathbfss{bold}{OT1}{cmss}{bx}{n} 
  \ifAMStwofonts 
    \ifCUPmtlplainloaded \else 
      \DeclareSymbolFont{UPM}{U}{eur}{m}{n} 
      \SetSymbolFont{UPM}{bold}{U}{eur}{b}{n} 
      \DeclareSymbolFont{AMSa}{U}{msa}{m}{n} 
      \DeclareMathSymbol{\upi}{0}{UPM}{"19} 
      \DeclareMathSymbol{\umu}{0}{UPM}{"16} 
      \DeclareMathSymbol{\upartial}{0}{UPM}{"40} 
      \DeclareMathSymbol{\leqslant}{3}{AMSa}{"36} 
      \DeclareMathSymbol{\geqslant}{3}{AMSa}{"3E} 

       \let\le=\leqslant 
       \let\ge=\geqslant 
    \fi 
  \fi 
\fi 
 
\ifCUPmtlplainloaded \else 
  \ifAMStwofonts \else 
    \def\upi{\pi} 
    \def\umu{\mu} 
    \def\upartial{\partial} 
  \fi 
\fi

\title{Global metallicity of globular cluster stars from  
colour-magnitude diagrams} 
\author[F. Caputo and S. Cassisi] 
       {F. Caputo$^1$ and S. Cassisi$^{2,3}$ 
\\ 
        $^1$ INAF-Osservatorio Astronomico di Roma, via di Frascati 33, 00040 Monte 
             Porzio Catone, Italy~~caputo@coma.mporzio.astro.it\\ 
        $^2$ INAF-Osservatorio Astronomico di Teramo, via M. Maggini, 64100 Teramo,  
             Italy~~cassisi@te.astro.it\\ 
        $^3$ Max-Planck fur Astrophysik, Karl-Schwarzschild-Strasse 1, 85741 Garching, Germany} 
\date{} 
 
\begin{document} 
 
\maketitle 
 
\label{firstpage} 
 
\begin{abstract} 
 
We have developed an homogeneous evolutionary scenario  
for H- and He-burning low-mass stars  
by computing updated stellar models for a   
wide metallicity and age range  
(0.0002$\le Z \le$0.004 and 9$\le t(Gyr) \le$15, respectively)  
suitable to study globular clusters.  
This theoretical scenario allows us to provide  
self-consistent predictions about the dependence 
of selected observational features of the  
colour-magnitude diagram, such as the brightness 
of the Turn Off (TO), Zero Age Horizontal Branch (ZAHB) and  
Red Giant Branch bump (BUMP), on the cluster 
metallicity and age. 
 
Taking into account these predictions,  
we introduce a new observable based on the visual magnitude  
difference between the TO and the ZAHB [$\Delta M_V$(TO-ZAHB)],  
and the TO and the RGB-bump  
[$\Delta M_V$(TO-BUMP)], given by  
$A=\Delta M_V$(TO-BUMP)$-$0.566$\Delta M_V$(TO-ZAHB).  
We show that the parameter $A$ 
does not depend at all on the cluster age, whereas it does strongly 
depend on the   
cluster global metallicity.  The calibration of the parameter 
$A$ as a function of $Z$ is then provided, as based  
on our evolutionary models.  
We tested the reliability of this result by also 
considering stellar models computed by other authors,
employing different input physics.
Eventually, we  
present clear evidence that the variation of  
$\Delta M_V$(TO-BUMP) with $\Delta M_V$(TO-ZAHB) does supply a  
powerful probe of the global metal abundance, at least when homogeneous  
theoretical frameworks are  adopted. Specifically, we show that     
the extensive set of models by VandenBerg et al. (2000) suggests  
a slightly different calibration of
$A$ versus $Z$ calibration, which however provides 
global metallicities higher by only 0.08$\pm$0.06 dex with respect  
to the results from our computations.    
 
We provide an estimate of the global metallicity of
36 globular clusters in the Milky Way,
based on our {\it A-Z} calibration, and a large  
observational database of Galactic globular clusters. 
By considering the empirical [Fe/H]  
scales by both Zinn \& West (1984) and 
Carretta \& Gratton (1997), we are able to provide an estimate of the 
$\alpha-$element enhancement for all clusters in our sample.  
We show that the trend of [$\alpha$/Fe] 
with respect to the iron 
content significantly depends on the  
adopted empirical [Fe/H] scale, with  
the Zinn \& West (1984) one suggesting  
$\alpha-$element enhancements in fine agreement with 
current spectroscopical measurements. 
 
\end{abstract} 
 

\section{Introduction} 
 
The colour-magnitude diagrams (CMDs) of  
globular clusters (GCs) present several features to be  
compared with the   
various constraints provided by stellar evolution theory.  
For this reason, they are often used to infer fundamental properties of   
ancient stellar populations and, in turn, of the early Universe.  
 
It has been known for a long time that the absolute magnitude of  
the main-sequence  
turnoff (TO) can be calibrated in terms of the cluster age ($t$, in Gyr)  
and chemical composition parameters ($Y$, helium content and  
$Z$, global metallicity).  
In  order to overcome the uncertainties of obtaining accurate GC  
distances, it is common to use  
the difference in magnitude between the TO and the  
zero-age-horizontal-branch (ZAHB) at the RR Lyrae instability strip  
(log$T_e\sim$3.85),  
whose luminosity  
level depends on  $Y$ and $Z$, with a negligible dependence on $t$. Besides  
this {\it vertical} method to get the cluster age, there is the so-called  
{\it horizontal} method, which is based on the 
determination of the TO colour relative  
to some fixed point on the red giant branch (RGB -- 
see Stetson, Vandenberg \& Bolte 1996).   
As a whole, both methods are affected by the empirical  
difficulties to get the precise  
position of the TO in the CMD,  
as well as by several still-open intrinsic uncertainties on the  
evolutionary models, most notably the dependence on $Z$  of   
the ZAHB luminosity, for the {\it vertical} method,  
and the efficiency of superadiabatic 
convection in the outer layers (i.e., the mixing length calibration)  
for the {\it horizontal} method, (see Rosenberg et al. 1999, hereafter R99,  
for details and references).  
 
The determination of $Z$ for GC stars is by itself problematic.  
Under the classical assumption of scaled-solar metal distribution,  
where all of the abundance ratios of the various metals relative to iron have  
the solar value, the correlation between the global metallicity $Z$  
and the measured iron-to-hydrogen ratio [Fe/H] is given by  
the classical relation log$Z$=[Fe/H]$-$1.70. One should however consider that current empirical  
scales for [Fe/H] (e.g., Zinn \& West 1984, Zinn 1985, Rutledge, Hesser \& Stetson 1997,  
and Carretta \& Gratton 1997) may differ by up to $\sim$0.3 dex.  
In addition, it has to be noticed that the assumption of scaled-solar chemical compositions  
might be rather inadequate for Galactic GC stars which show a not negligible overabundance of  
the so-called $\alpha-$elements ( O, Ne, Mg, Si, Ar, Ca, and Ti)  
with respect to  
iron-peak elements ([$\alpha$/Fe]$\sim$ 0.3, according to Carney 1996).  
It follows that the relationship
between the measured [Fe/H] and the overall metallicity $Z$ should follow  
the relation  
$$\log Z=[Fe/H] -1.70+\log (0.638f+0.362)\eqno(1)$$ 
\noindent 
where $f$ is the enhancement factor of $\alpha-$elements with respect to iron  
(Salaris et al. 1993)\footnote{It is worth recalling that this
relation holds strictly only in case of 
the scaled-solar metal mixture by Ross \& Aller (1976). The coefficients of this relation has to be 
slightly changed when accounting for different solar metal distributions. However, these changes are 
very small (see Yi et al. 2001).}.   
 
The knowledge of the cluster age and global metallicity is fundamental for studying another  
important feature of CMDs, i.e. the luminosity level of the so-called RGB-bump.  
This is  an intriguing feature of the RGB luminosity function which has been predicted for a long time,  
since the pioneering theoretical studies by Thomas (1967) and Iben (1968), but whose identification  
in GCs occurred years later (King et al. 1985). A first comparison
(Fusi Pecci et al. 1990) of the
observed   
difference in visual magnitude between  RGB-bump and  ZAHB level with theoretical models  
showed that the predicted dependence of $\Delta V$(ZAHB-BUMP) on metallicity was nicely reproduced,  
but the zero point of the observed relation was $\sim$ 0.4 mag too faint.  
 
An exhaustive investigation on the dependence of the parameter $\Delta V$(ZAHB-BUMP) 
on the physical inputs adopted in stellar computations has been performed by Cassisi \& Salaris (1997, 
hereinafter CS97). 
In the same work, it was clearly shown that a fine agreement between theory and observations 
does exist
when updated RGB stellar models and observational data for GCs with
spectroscopic 
measurements of both iron and $\alpha-$element abundances, are considered. 
More recently, the same result was obtained by Zoccali et al. (1999) and Ferraro et al. (1999, 
hereinafter F99) by using a much larger sample of Galactic GCs than CS97.  
However, one has also to notice that 
Bergbusch \& Vandenberg~(2001) have recently found a discrepancy of the order of 0.25 mag, when comparing 
the $\Delta V$(ZAHB-BUMP) values provided by their models with the
observational data for 4 GCs 
at various metallicities. 
 
In reality, the RGB-bump luminosity is also dependent on the age of the 
stellar population ($\delta V$(BUMP)/$\delta$log$t\sim$ 1), while the above investigations do not take  
into account  the age of individual GCs. It is well known 
the current debate whether the Galactic GCs formed all at the same
time or there was a significantly protracted formation epoch. 
A recent analysis of an homogeneous photometric database containing 34  
Galactic GCs (R99) shows that within the intermediate-metallicity clusters ($-1.2\le$[Fe/H]$\le-$0.9)  
there is a clear evidence of age dispersion, with some clusters up to $\sim$30\% younger 
than the oldest ones. As an example, we consider 
the globular cluster NGC 362 for which F99 give $\Delta V$(ZAHB-BUMP)=0.10 mag$\pm$0.12 mag and a  
global metallicity\footnote{[M/H]=[Fe/H]+log(0.638$f$+0.362).} [M/H]=$-$0.99.  
 
{From} the relations given by these authors  
$$M_{V,BUMP}=0.75+0.99\log t+1.58[M/H]+0.26[M/H]^2\eqno(2)$$ 
$$M_{V,ZAHB}=1.00+0.35[M/H]+0.05[M/H]^2\eqno(3)$$ 
\noindent 
one has that the agreement between theory and observations occurs for  
a cluster age of $\sim$14.9 Gyr, which is the value derived by R99 for the bulk of coeval GCs,   
using the Straniero et al. (1997) theoretical isochrones. However, R99 find that the actual age  
of NGC 362 is $\sim$ 11.5 Gyr, in which case one obtains a $\Delta
V$(ZAHB-BUMP) value by 0.21 mag, namely $\sim$0.1 mag larger than the observed one.   
 
It should be noticed that R99 adopts the [Fe/H] scale by Carretta \& Gratton (1997), 
without any correction due to $\alpha-$enhancement and, in order to interpret the observed   
difference $\Delta V$(TO-ZAHB), a linear relation for the absolute magnitude of the ZAHB as a  
function of the metal content ($M_{V,ZAHB}$=0.18([Fe/H]+1.5)+0.65), instead of  
the quadratic relations used by F99 [see Eq. (3)] and CS97   
($M_{V,ZAHB}$=1.13+0.39[M/H]+0.06[M/H]$^2$).  
In other words, the quoted extensive investigations on the RGB-bump luminosity and  
the age of Galactic GCs are not mutually consistent, since they use different metallicity scales and  
adopt, as theoretical values for the RGB-bump and TO luminosity,    
different calibrations of the ZAHB luminosity in terms of metallicity.  
 
In order to make a complete and self-consistent study of these most prominent 
features in  CMDs, we present in section 2 an homogeneous set of theoretical  
evolutionary results which predict the TO, RGB-bump and ZAHB  
luminosity for a wide range of ages  
(9$\le{t(Gyr)}\le$15) and  metal abundances (0.0002$\le{Z}\le$0.004).  
In section 3 we introduce a new parameter $A$ defined as  
$A=\Delta M_V$(TO-BUMP)$-$0.566$\Delta M_V$(TO-ZAHB), and we 
show that this depends only on the global metallicity $Z$,  
being unaffected by age.  
In order to test this result, the evolutionary models computed by  
Salaris \& Weiss (1997, 1998) and by Vandenberg et al. (2000) are also considered. We show that  
the former ones are in close agreement with our computations, whereas  
in the latter case some significant differences are present, leading to   
a slightly different formulation of the  
$\Delta M_V$(TO-BUMP) versus $\Delta M_V$(TO-ZAHB) relationship.  
Nevertheless, the parameter   
$A$=$\Delta M_V$(TO-BUMP)$-$0.487$\Delta M_V$(TO-ZAHB) inferred from  
the Vandenberg et al.'s (2000) computations is again strongly dependent on global metal abundance with  
no dependence on age.  
     
On this basis, the TO, ZAHB and RGB-bump observations collected by F99 and R99  
for a large sample of Galactic GCs are interpreted. The resulting  
global metal abundances are  
presented in section 4, together with some clues to  
the $\alpha-$element enhancement with respect to iron in Galactic GCs.  
A summary is finally given in the last section. 
 
\begin{table} 
\centering 
\caption[]{Selected parameters from the evolutionary models.\label{tab1}} 
\begin{tabular}{lrccc} 
          $Z$   &  Age(Gyrs)  & $M_V$(TO) & $M_V$(Bump) & $M_V$(ZAHB) \\ 
 
           0.0002  &    9  &  3.543 &  -0.394 &   0.443 \\ 
           0.0002  &   10  &  3.677 &  -0.339 &   0.439 \\ 
           0.0002  &   11  &  3.794 &  -0.288 &   0.436 \\ 
           0.0002  &   12  &  3.923 &  -0.241 &   0.433 \\ 
           0.0002  &   13  &  4.010 &  -0.199 &   0.430 \\ 
           0.0002  &   14  &  4.090 &  -0.161 &   0.427 \\ 
           0.0002  &   15  &  4.164 &  -0.127 &   0.425 \\ 
           0.0006  &    9  &  3.733 &  -0.016 &   0.533 \\ 
           0.0006  &   10  &  3.855 &   0.036 &   0.529 \\ 
           0.0006  &   11  &  3.945 &   0.080 &   0.526 \\ 
           0.0006  &   12  &  4.067 &   0.109 &   0.523 \\ 
           0.0006  &   13  &  4.178 &   0.152 &   0.520 \\ 
           0.0006  &   14  &  4.247 &   0.192 &   0.517 \\ 
           0.0006  &   15  &  4.312 &   0.227 &   0.515 \\ 
           0.001   &    9  &  3.817 &   0.158 &   0.563 \\ 
           0.001   &   10  &  3.926 &   0.218 &   0.560 \\ 
           0.001   &   11  &  4.023 &   0.253 &   0.557 \\ 
           0.001   &   12  &  4.184 &   0.290 &   0.555 \\ 
           0.001   &   13  &  4.256 &   0.318 &   0.552 \\ 
           0.001   &   14  &  4.323 &   0.370 &   0.550 \\ 
           0.001   &   13  &  4.256 &   0.318 &   0.552 \\ 
           0.001   &   14  &  4.323 &   0.370 &   0.550 \\ 
           0.001   &   15  &  4.386 &   0.425 &   0.548 \\ 
           0.002   &    9  &  3.982 &   0.525 &   0.584 \\ 
           0.002   &   10  &  4.083 &   0.571 &   0.581 \\ 
           0.002   &   11  &  4.209 &   0.610 &   0.578 \\ 
           0.002   &   12  &  4.284 &   0.642 &   0.576 \\ 
           0.002   &   13  &  4.362 &   0.691 &   0.574 \\ 
           0.002   &   14  &  4.429 &   0.739 &   0.572 \\ 
           0.002   &   15  &  4.491 &   0.770 &   0.570 \\ 
           0.003   &    9  &  4.066 &   0.753 &   0.665 \\ 
           0.003   &   10  &  4.157 &   0.778 &   0.663 \\ 
           0.003   &   11  &  4.239 &   0.848 &   0.660 \\ 
           0.003   &   12  &  4.357 &   0.890 &   0.658 \\ 
           0.003   &   13  &  4.425 &   0.912 &   0.656 \\ 
           0.003   &   14  &  4.464 &   0.933 &   0.654 \\ 
           0.003   &   15  &  4.571 &   0.952 &   0.652 \\ 
           0.004   &    9  &  4.023 &   0.902 &   0.714 \\ 
           0.004   &   10  &  4.165 &   0.961 &   0.711 \\ 
           0.004   &   11  &  4.241 &   1.014 &   0.709 \\ 
           0.004   &   12  &  4.354 &   1.067 &   0.707 \\ 
           0.004   &   13  &  4.414 &   1.102 &   0.704 \\ 
           0.004   &   14  &  4.501 &   1.128 &   0.703 \\ 
           0.004   &   15  &  4.562 &   1.135 &   0.701 \\ 
\end{tabular} 
\end{table}

\begin{figure} 
\psfig{figure=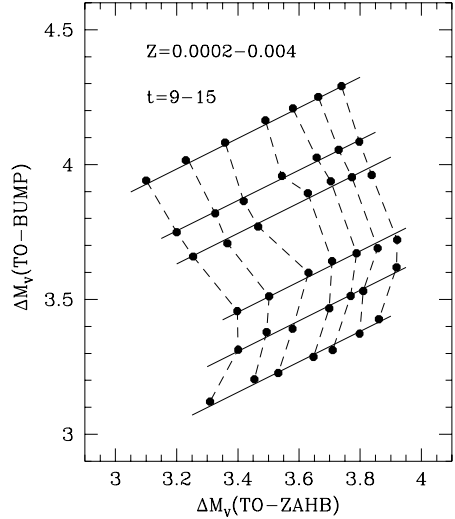,height=10cm} 
\caption{Theoretical predictions for $\Delta M_V$(TO-BUMP) 
and $\Delta M_V$(TO-ZAHB) as a function of age ($t$) and 
global metallicity ($Z$). The dashed 
lines refer to constant ages in the range of 
9 to 15 Gyr (left to right), while the 
solid lines refer to constant metallicities in the range of 
0.0002 to 0.004 (bottommost line). The solid lines have 
a similar slope of 0.566 (see text).} 
\end{figure} 
 
\begin{figure} 
\psfig{figure=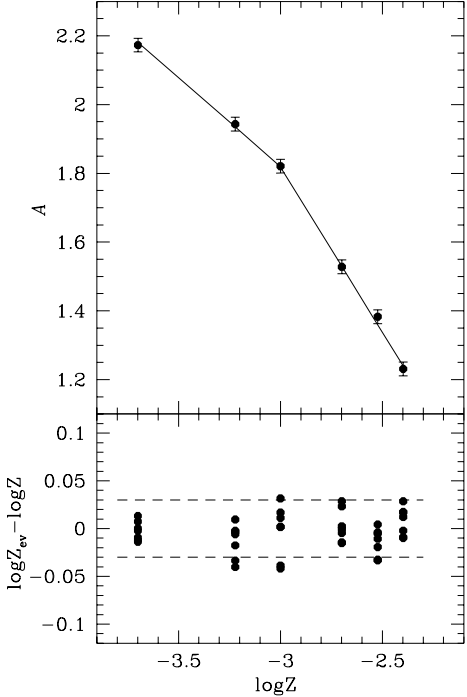,height=10cm} 
\caption{({\it top}) - Mean value of the parameter 
$A$=$\Delta M_V$(TO-BUMP)$-$0.566$\Delta M_V$(TO-ZAHB) as a function of $Z$. 
The solid line is the analytic fit to the data 
(see Eq. (4) and Eq. (5) in the text). ({\it bottom}) -  
Difference between the {\it evolutionary} metallicity 
log$Z_{ev}$ derived using Eq. (4) and Eq. (5) and the original 
value log$Z$, as a function of log$Z$. 
The discrepancy is within $\pm$0.03 dex (dashed line).} 
\end{figure}

\section{Theoretical models} 
 
In order to understand the observed properties of CMDs and RGB luminosity 
functions of GCs, we need an homogeneous set of evolutionary models 
for both the H- and He-burning phases. In present work, we have adopted 
a set of theoretical models computed by using the FRANEC evolutionary 
code (Chieffi \& Straniero 1989; 
CS97; Castellani et al. 1997). 
A subset of these stellar models and corresponding isochrones have been already 
presented by Cassisi et al. (1998, 1999). However, in order to 
improve the coverage of the wide metallicity range spanned by Galactic GCs, 
we have extended previous 
computations considering additional initial chemical compositions. 
The physical inputs adopted in the model computations have been 
already discussed by Cassisi et al. (1998). A short summary is given below.
 
We use the OPAL (Rogers et al. 1996)
equation of state (EOS), supplemented, in the temperature-density regime 
not covered by the OPAL tables, by the EOS provided by Straniero (1988), 
plus a Saha EOS in the outer stellar layers. The radiative opacities  
provided by the OPAL group (Iglesias, Rogers \& Wilson 1992) for temperatures larger 
than 10,000K are employed, together with the
Alexander \& Ferguson (1994) tables for lower temperatures. 
Electron conduction opacities come from Itoh et al. (1983). 
The boundaries of the convective regions are determined according to the  
canonical Schwarzschild criterion. Induced overshooting and semiconvection  
during the central He-burning phase have been accounted following the prescriptions  
given by Castellani et al. (1985). The standard mixing-length theory is used for evaluating  
the temperature gradient in the superadiabatic regions; the value of the mixing length  
parameter is fixed at 1.6 for all computations. 
This value allows to match the RGB effective temperatures 
of a selected sample of Galactic  
GCs provided  by Frogel, Persson \& Cohen (1983). 
 
All the models include microscopic diffusion of both  
helium and heavy elements, using the formalism by Thoul et al. (1994). 
For a detailed discussion concerning the effects of atomic diffusion 
on both  main-sequence stars and more evolved models, we refer 
the interested reader to Castellani et al. (1997) and Cassisi et al. (1998). 
It is well known that helioseismology provides clear evidence that helium and  
heavy element settling is occurring in the Sun (see Basu, Pinsonneault, \& Bahcall 2000,  
and references therein).  
However, recent spectroscopic analyses of GC stars  
(Gratton et al. 2001; Castilho et al. 2000;  Ramirez et al. 2001)  
show that the
surface chemical patterns do not show the effects of atomic diffusion,  
at least for what heavy elements are concerned.  
This means that either 
chemical settling is less efficient in metal-poor stars, or some other physical 
mechanism at work in the stellar atmospheres counterbalances the effect 
of atomic diffusion. A firmer understanding of this topic has to wait for 
more accurate theoretical and observational investigations.   
 
Evolutionary models with masses  
ranging from 0.6 to 1.3$M_\odot$ have been computed from the  
Zero Age  Main Sequence up to the tip of the RGB marking the  
triple-alpha ignition inside the degenerate He core. The explored range 
of metallicity is $0.0002\le{Z}\le0.004$, while for each metallicity the same initial  
He abundance $Y$=0.23 has been adopted. 
The He-core mass $M_c$ and the chemical stratification at the  
RGB He-flash provided by the different 
evolutionary models have been used   
to compute the ZAHB structures needed to estimate the ZAHB luminosity level 
at log$T_e$=3.85, the average effective temperature of the RR Lyrae instability strip.  
Isochrones have been then constructed in the age range 9 - 15 Gyr.
 
Finally, all the models and isochrones have been transposed from the theoretical H-R plane to 
the observational ($V, B-V$) plane by using the bolometric corrections and 
colour-temperature relations provided by Castelli, Gratton \& Kurucz (1997a,b). 
In Table 1, for the different chemical compositions adopted in  
present work, we report the TO, RGB-bump and ZAHB absolute visual  
magnitude, as a function of age.

\section{Theoretical calibrations}  
 
The data listed in Table 1 show that, at a fixed age, TO, RGB-bump and ZAHB  
luminosities decrease with increasing the metal content. At a fixed metallicity  
the TO and RGB-bump become fainter for larger ages, whereas the ZAHB level becomes  
slightly brighter, due to the slight increase of the He-core mass $M_c$ with decreasing  
total mass of the progenitor on the RGB.  
 
Figure 1 displays the two differential parameters $\Delta M_V$(TO-ZAHB) and  
$\Delta M_V$(TO-BUMP), as derived from the results in Table 1.  
The dashed lines show the theoretical trend at a constant age, in
steps of 1 Gyr, from 9 Gyr (left) to 15 Gyr (right), while the solid lines refer to a  
constant $Z$ from 0.0002 to 0.004 (bottommost line).  
The difference $\Delta M_V$(TO-BUMP) increases for larger ages,  
at fixed $Z$, and decreases for larger metallicities, at fixed $t$. As for the parameter  
$\Delta M_V$(TO-ZAHB), it increases with increasing $Z$ up to log$Z\sim-$2.7, and then it  
decreases for larger metallicities, at least when $Z\le$0.004. On this  
subject, we recall that all our models are computed with an initial helium abundance    
$Y$=0.23. Within the quoted range of metal content, the expected He increase  
due to an helium-to-heavy element enrichment ratio $\Delta Y/\Delta Z\sim$2.5 is only  
$\Delta Y$=0.01, whereas for metallicities larger than $Z$=0.004 the original amount  
of helium is expected to become significantly larger. Furthermore, also the amount  
of extra-helium brought to the stellar surface by the first dredge up is expected to increase  
when moving towards larger metallicities. As a consequence, the above theoretical trends  
{\em cannot} be applied to very metal-rich clusters ([Fe/H]$\ge-0.6$).     
 
The data in Fig. 1 disclose that, for ages between 9 and 15 Gyr,  
the variation of $\Delta M_V$(TO-BUMP) with $\Delta M_V$(TO-ZAHB)  
has a constant slope of 0.566 (solid lines), with the zero-point depending on $Z$.    
From the results listed in Table 1 we calculate then the mean value of the  
parameter $A$=$\Delta M_V$(TO-BUMP)$-$0.566$\Delta M_V$(TO-ZAHB) for each given $Z$.  
The upper diagram in  Fig. 2 shows that the results are well approximated by the  
following linear relations:  
 
$$\log Z=+0.518-1.933A\eqno(4)$$ 
$$\log Z=-1.110-1.039A\eqno(5)$$    
\noindent 
with $A\ge 1.82$ and $A<1.82$, respectively. As it is shown in the lower panel  
of Fig. 2,   
the discrepancy between the {\it evolutionary} estimate of the  
global metallicity (hereafter $Z_{ev}$) based on the two differential  
parameters $\Delta M_V$(TO-BUMP) and $\Delta M_V$(TO-ZAHB), and the original  
value, is close to $\pm$0.03 dex.

\begin{figure} 
\psfig{figure=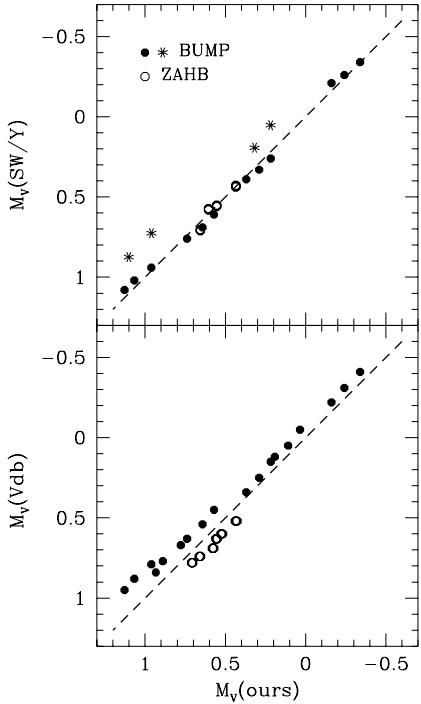,height=10cm} 
\caption{({\it top}) - Comparison between our theoretical predictions for the RGB Bump and 
ZAHB luminosity and the ones provided by Salaris \& Weiss  
(1997, 1998: open and filled circles, respectively) and 
Yi et al. (2001: asterisks). Each point corresponds to a fixed age and metallicity,  
for the values in common with our models. 
{\it bottom}) - As top panel, but for the models
by Vandenberg et al. (2000). Open circles refer to the ZAHB luminosity, while full  
circles correspond to the RGB Bump luminosity.} 
\end{figure} 
 
\begin{figure} 
\psfig{figure=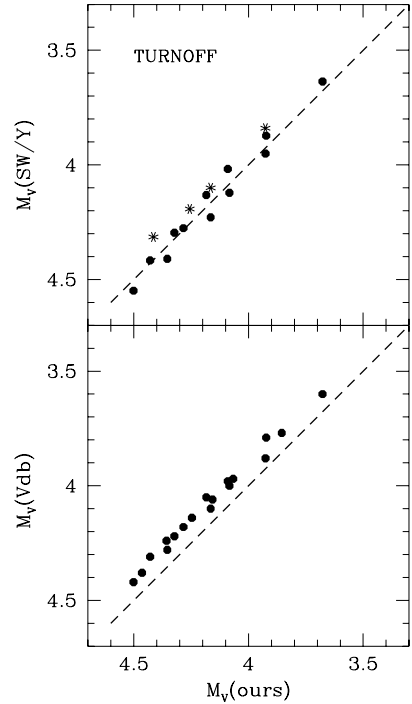,height=10cm} 
\caption{As in Fig. 3, but for the TO luminosity.} 
\end{figure} 
 
\begin{figure} 
\psfig{figure=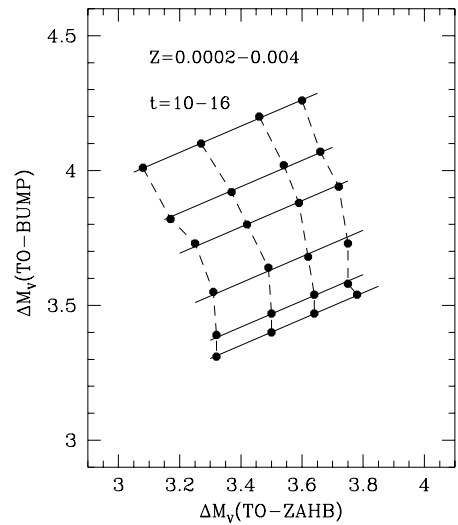,height=10cm} 
\caption{As in Fig. 1, but in this case the Vdb models have been adopted.} 
\end{figure} 
 
\begin{figure} 
\psfig{figure=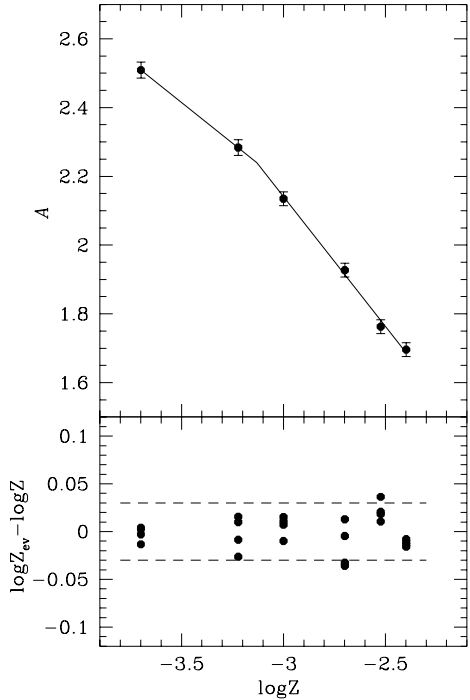,height=10cm} 
\caption{As in Fig. 2, but adopting the Vdb models.} 
\end{figure} 
 
Before proceeding further, it is worthwhile to compare our evolutionary 
predictions with similar computations provided by different authors. 
We considered the recent isochrones and ZAHB models computed by 
Salaris \& Weiss (1997, 1998, hereinafter SW) and by Vandenberg et al. (2000, hereinafter Vdb, 
see also Bergbusch \& Vandenberg 2001) with $Z$ ranging from 0.0002 to 0.004  
and ages in the range of 10-14 Gyr and 10-16 Gyr, respectively. 
We wish to notice that both sets of models do not account for  
atomic diffusion and have been computed by adopting a slightly different $\Delta Y/\Delta Z$  
ratio (3 and 2, respectively). Moreover, they are computed  
under different assumptions about the $\alpha-$element enhancement.  
In our comparison we used the Vdb models based on a scaled-solar heavy element distribution.  
 
As shown in the upper panel of Fig. 3, both the ZAHB (open circles) and  
RGB-bump (filled circles)  
absolute magnitudes inferred from   
SW models are in close agreement with our results, whereas the  
Vdb computations  
plotted in the lower panel suggest brighter RGB-bump and 
fainter ZAHB luminosities, with respect to the data in Table 1.  
The upper panel of Fig. 3 shows also the comparison between  
the RGB-bump absolute magnitudes provided by the recent Yale isochrones  
(Yi et al. 2001) and our results (asterisks),  
for the common values of metallicity  
($Z$=0.001 and 0.004) and age (10 and 13 Gyrs). As a whole, one has that  
the Yale isochrones suggest brighter RGB-bumps than our models. Note that  
we were unable to carry out a similar comparison with  
the Yale ZAHB absolute visual magnitudes, as no data  
have been provided in the literature.   
Figure 4 deals with the TO luminosity provided by the quoted sets of  
computations,   
compared with the results in Table 1. We get that SW models are again  
in agreement with our predictions, whereas VdB and Yale models give  
brighter absolute magnitudes.  
 
On the basis of the above discussion, we have considered  
the models by VdB for testing the effect of  
different input physics on the scenario suggested by our computations.   
Figure 5 presents $\Delta M_V$(TO-BUMP) versus $\Delta M_V$(TO-ZAHB),  
as derived from Vdb models. The dashed lines show the theoretical trend at  
a constant age, in steps of 2 Gyr, from 10 Gyr (left) to 16 Gyr (right),  
while the solid lines refer to a constant $Z$ from 0.0002 to 0.004 (bottommost line).  
The general trend in Fig. 1 is here confirmed, but the variation of  
$\Delta M_V$(TO-BUMP) with $\Delta M_V$(TO-ZAHB) has a slope of  
0.487 (solid lines). The correlation between the mean value of the  
parameter  
$A$=$\Delta M_V$(TO-BUMP)$-$0.487$\Delta M_V$(TO-ZAHB)  
and log$Z$ is presented in the   
upper diagram of Fig. 6, where the solid line  
depicts the relations  
 
$$\log Z=+1.617-2.119A\eqno(6)$$ 
$$\log Z=-0.154-1.330A\eqno(7)$$ 
\noindent 
with $A\ge 2.24$ and $A<2.24$, respectively. As shown in the lower panel 
of Fig. 6, also in this case  
the discrepancy between the {\it evolutionary} estimate of the 
global metallicity (hereafter $Z_{ev}$)  
and the original 
value is close to $\pm$0.03 dex.

\begin{table*} 
\centering 
\caption[]{Observed parameters for Galactic Globular Clusters.  
Iron-to-hydrogen contents are given   
in the Zinn \& West (1984, [Fe/H]$_{ZW}$) and Carretta \& Gratton (1997, [Fe/H]$_{CG}$)  
empirical scales. For the latter scale, the fourth column gives the source:  
0 = Carretta \& Gratton (1997);  1 = Rutledge, Hesser \& Stetson (1997);  
2 = Ferraro et al. (2000); 3 = Rosenberg et al. (1999). The [Fe/H]$_{CG}$ value of NGC 7492  
has been estimated through the mean relation between  
[Fe/H]$_{CG}$ and [Fe/H]$_{ZW}$. \label{tab2}} 
\begin{tabular}{rccccccccc} 
 
NGC & [Fe/H]$_{ZW}$ & [Fe/H]$_{CG}$ & Source & $V$(TO)  & $V$(ZAHB) & $V$(BUMP) \\ 
  104 & $-0.71$ & $-$0.70 & 0   & 17.60$\pm$0.08 &  14.22$\pm$0.07 & 14.55$\pm$0.05 \\ 
  288 & $-1.40$ & $-$1.07 & 0   & 18.90$\pm$0.04 &  15.50$\pm$0.10 & 15.45$\pm$0.05 \\ 
  362 & $-1.27$ & $-$1.15 & 0   & 18.79$\pm$0.04 &  15.50$\pm$0.07 & 15.40$\pm$0.10 \\ 
 1261 & $-1.29$ & $-$1.08 & 1,2 & 19.90$\pm$0.06 &  16.72$\pm$0.05 & 16.60$\pm$0.05 \\ 
 1851 & $-1.33$ & $-$1.05 & 1,2 & 19.50$\pm$0.07 &  16.20$\pm$0.05 & 16.15$\pm$0.05 \\ 
 1904 & $-1.68$ & $-$1.37 & 0   & 19.65$\pm$0.09 &  16.27$\pm$0.07 & 15.95$\pm$0.05 \\ 
 2808 & $-1.37$ & $-$1.13 & 1,2 & 19.60$\pm$0.07 &  16.27$\pm$0.07 & 16.15$\pm$0.05 \\ 
 3201 & $-1.56$ & $-$1.23 & 0   & 18.20$\pm$0.05 &  14.77$\pm$0.07 & 14.55$\pm$0.05 \\ 
 4590 & $-2.09$ & $-$1.99 & 0   & 19.05$\pm$0.07 &  15.75$\pm$0.05 & 15.15$\pm$0.05 \\ 
 4833 & $-1.86$ & $-$1.58 & 0   & 19.20$\pm$0.10 &  15.77$\pm$0.07 & 15.35$\pm$0.05 \\  
 5272 & $-1.66$ & $-$1.34 & 0   & 19.10$\pm$0.04 &  15.68$\pm$0.05 & 15.45$\pm$0.05 \\ 
 5286 & $-1.79$ & $-$1.49 & 1,2 & 20.05$\pm$0.10 &  16.60$\pm$0.10 & 16.25$\pm$0.05 \\  
 5694 & $-1.92$ & $-$1.73 & 1,2 & 22.12$\pm$0.10 &  18.70$\pm$0.10 & 18.15$\pm$0.07 \\  
 5897 & $-1.68$ & $-$1.59 & 0   & 19.75$\pm$0.07 &  16.45$\pm$0.07 & 16.00$\pm$0.10 \\ 
 5904 & $-1.40$ & $-$1.11 & 0   & 18.50$\pm$0.03 &  15.13$\pm$0.05 & 15.00$\pm$0.05 \\ 
 6093 & $-1.68$ & $-$1.44 & 1,2 & 19.80$\pm$0.08 &  16.12$\pm$0.07 & 15.95$\pm$0.10 \\ 
 6121 & $-1.28$ & $-$1.19 & 0   & 16.90$\pm$0.03 &  13.45$\pm$0.10 & 13.40$\pm$0.10 \\ 
 6171 & $-0.99$ & $-$0.95 & 1,2 & 19.25$\pm$0.06 &  15.70$\pm$0.10 & 15.85$\pm$0.05 \\ 
 6205 & $-1.65$ & $-$1.39 & 0   & 18.50$\pm$0.06 &  15.10$\pm$0.15 & 14.75$\pm$0.07 \\ 
 6218 & $-1.61$ & $-$1.26 & 1,2 & 18.30$\pm$0.07 &  14.75$\pm$0.15 & 14.60$\pm$0.07 \\ 
 6229 & $-1.54$ & $-$1.30 & 2   & 21.48$\pm$0.10 &  18.11$\pm$0.05 & 18.00$\pm$0.07 \\ 
 6254 & $-1.60$ & $-$1.41 & 0   & 18.55$\pm$0.05 &  14.85$\pm$0.10 & 14.65$\pm$0.05 \\ 
 6341 & $-2.24$ & $-$2.16 & 0   & 18.55$\pm$0.06 &  15.30$\pm$0.10 & 14.65$\pm$0.05 \\ 
 6362 & $-1.08$ & $-$0.96 & 0   & 18.90$\pm$0.08 &  15.41$\pm$0.06 & 15.60$\pm$0.02 \\ 
 6397 & $-1.91$ & $-$1.82 & 0   & 16.40$\pm$0.04 &  13.00$\pm$0.10 & 12.60$\pm$0.10 \\ 
 6584 & $-1.54$ & $-$1.30 & 2   & 20.00$\pm$0.10 &  16.60$\pm$0.05 & 16.40$\pm$0.10 \\ 
 6656 & $-1.75$ & $-$1.41 & 0   & 17.80$\pm$0.07 &  14.25$\pm$0.10 & 13.90$\pm$0.10 \\ 
 6681 & $-1.51$ & $-$1.31 & 1,2 & 19.25$\pm$0.09 &  15.85$\pm$0.10 & 15.65$\pm$0.05 \\ 
 6717 & $-1.32$ & $-$1.09 & 1,2 & 19.25$\pm$0.10 &  15.75$\pm$0.15 & 15.75$\pm$0.10 \\ 
 6752 & $-1.54$ & $-$1.42 & 0   & 17.35$\pm$0.08 &  13.90$\pm$0.15 & 13.65$\pm$0.05 \\ 
 6809 & $-1.82$ & $-$1.58 & 1,2 & 17.95$\pm$0.12 &  14.60$\pm$0.10 & 14.15$\pm$0.05 \\ 
 6838 & $-0.58$ & $-$0.70 & 0   & 17.95$\pm$0.06 &  14.52$\pm$0.10 & 14.80$\pm$0.15 \\ 
 6934 & $-1.54$ & $-$1.30 & 2   & 20.40$\pm$0.15 &  17.05$\pm$0.04 & 16.78$\pm$0.10 \\ 
 7006 & $-1.59$ & $-$1.35 & 2   & 22.30$\pm$0.10 &  18.85$\pm$0.15 & 18.55$\pm$0.07 \\ 
 7078 & $-2.15$ & $-$2.12 & 0   & 19.25$\pm$0.06 &  15.90$\pm$0.07 & 15.25$\pm$0.05 \\ 
 7492 & $-1.51$ & $-$1.38 &     & 21.25$\pm$0.10 &  17.78$\pm$0.10 & 17.55$\pm$0.10 \\ 
 
\end{tabular} 
\end{table*} 
 
\section{Global metallicity for  
globular clusters in the Galaxy} 
 
In this section we provide accurate {\it evolutionary}  
estimates of the global metallicity 
of a large sample of Galactic GCs, by using the relationships 
discussed in the previous section.

\begin{figure} 
\psfig{figure=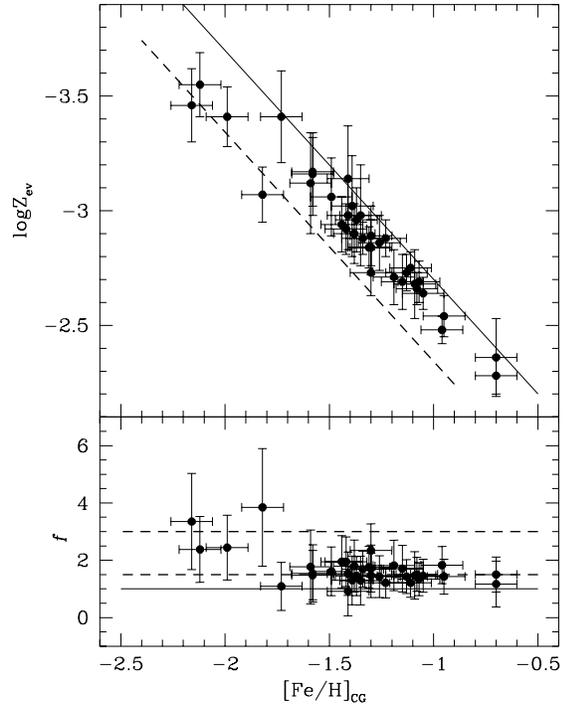,height=10cm} 
\caption{({\it top}) - The {\it evolutionary} global metallicity  
log$Z_{ev}$ for the GCs
in Table 3, as a function of  
the measured [Fe/H] value (with an uncertainty of $\pm$ 0.10 dex)  
on the Carretta \& Gratton (1997) scale. The solid line represents  
the canonical correlation for scaled-solar chemical compositions  
[$f$=1 in Eq. (1)], while the dashed line refers to $f$=3.  
({\it bottom}) - The $\alpha-$enhancement factor $f$ for the 
clusters in Table 3 
as a function of the measured [Fe/H] value 
on the Carretta \& Gratton (1997) scale. The solid line represents 
the classical value for scaled-solar chemical compositions 
($f$=1), while the dashed lines denote the average values 
$f\sim$ 1.5 and $f\sim$ 3 suggested by clusters with [Fe/H]$\ge -$1.7 and  
[Fe/H]$<-$1.7, respectively.} 
\end{figure} 
 
\begin{figure} 
\psfig{figure=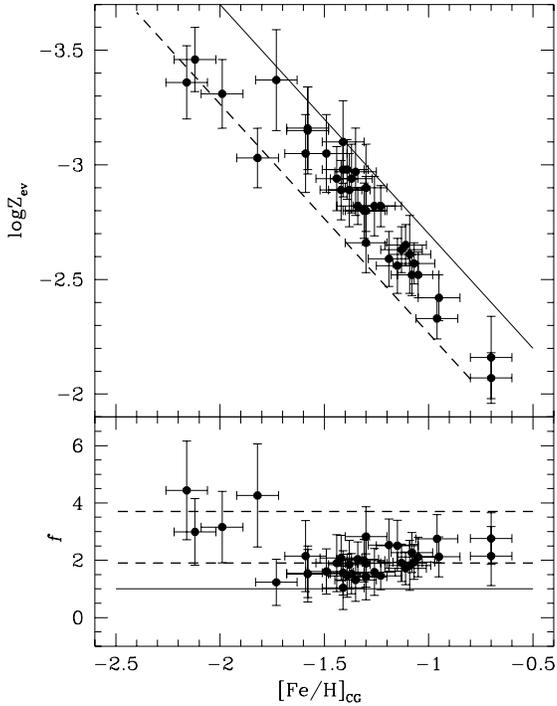,height=10cm} 
\caption{As in Fig. 7, but for Vdb models. The dashed line in the top  
panel refers to $f$=3.7, while the dashed lines in the bottom panel  
denote the average values 
$f\sim$ 1.9 and $f\sim$ 3.7 suggested by clusters with [Fe/H]$\ge -$1.7 and 
[Fe/H]$<-$1.7, respectively.} 
\end{figure} 
 
\begin{figure} 
\psfig{figure=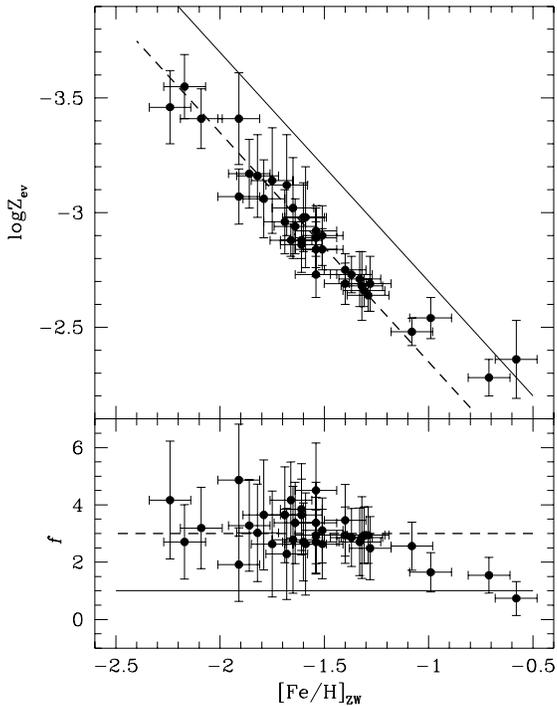,height=10cm} 
\caption{As in Fig. 7, but for the [Fe/H] scale by Zinn \& West (1984). 
The dashed line in both panels refer to an $\alpha-$enhancement  
factor $f\sim$ 3 suggested  
by clusters with [Fe/H]$\le-$1.3.} 
\end{figure} 
 
\begin{figure} 
\psfig{figure=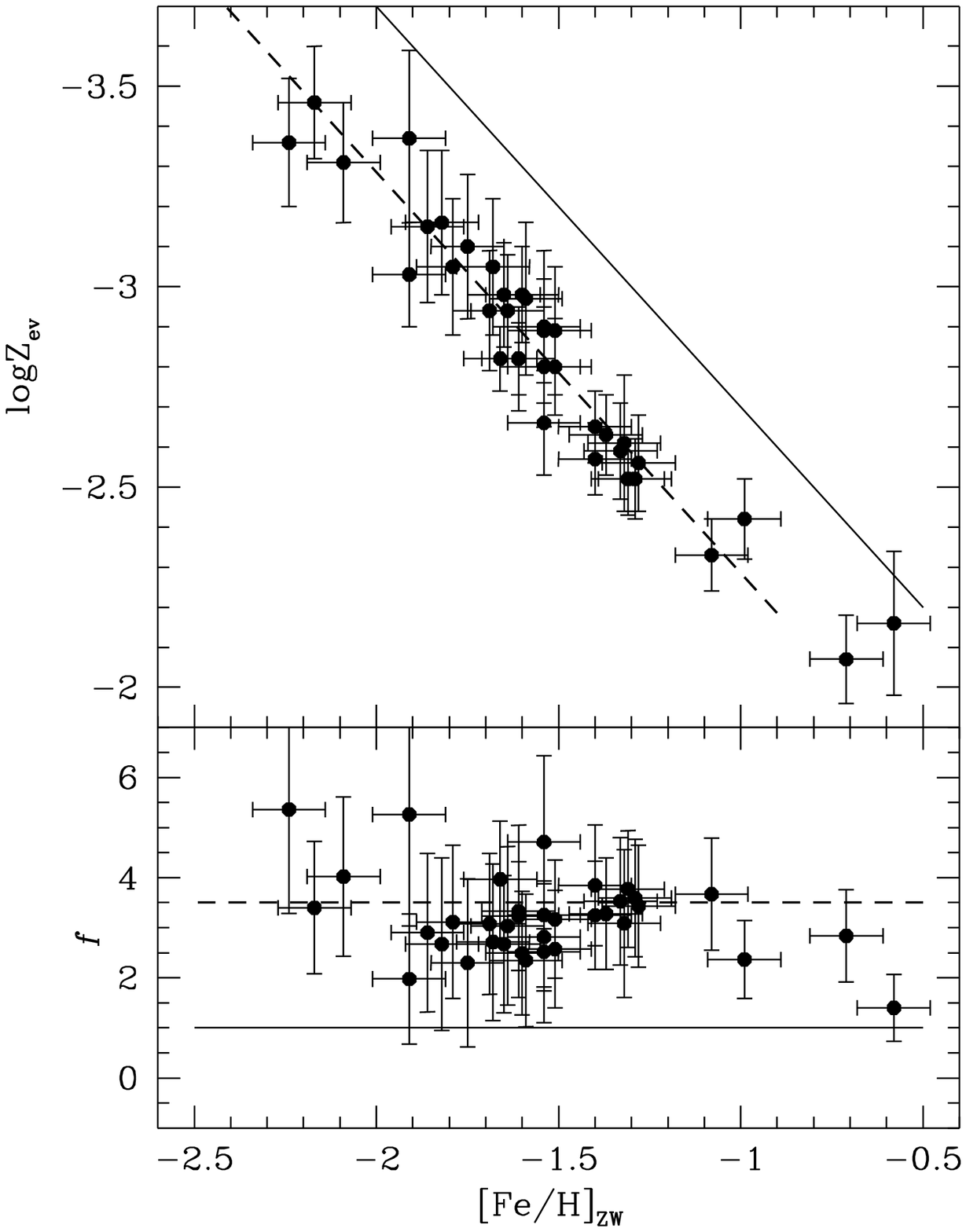,height=10cm} 
\caption{As in Fig. 9, but for Vdb models. The dashed line in both panels 
refers to an $\alpha-$enhancement factor $f\sim$ 3.5 suggested 
by clusters with [Fe/H]$\le-$1.3} 
\end{figure}

The sample of Galactic GCs studied in this paper is listed in Table 2.  
The [Fe/H] values are listed both for the Zinn \& West (1984) and the Carretta \& Gratton (1997)  
metallicity scale, while the TO visual magnitudes are from R99, except
for the  following clusters:   
NGC 4833 and 5286 (Samus et al. 1995a,b), NGC 5694  
(Ortolani \& Gratton 1990), NGC 6584 (Sarajedini \& Forrester 1995),  
NGC 6717 (Ortolani et al. 1999), NGC 6934 (Piotto et al. 1999)  
NGC 7006 (Buonanno et al. 1991) and  
NGC 7492 (Buonanno et al. 1987). The    
ZAHB and RGB-bump observations are from F99, except for the following  
additional measurements of the RGB-bump level: NGC 6352 and  6397  
(Alves \& Sarajedini 1999) and NGC 6362 (Piotto et al. 1999).  
We wish to notice that the two major  
sets of observations are mutually consistent  
as the ZAHB visual magnitudes listed by F99 are $\sim$ 0.07 mag fainter  
than the HB visual magnitudes in R99, except for NGC 6093 and NGC 6254.  
For these two clusters, the F99 values are {\it brighter} than those given  
by R99 (see later). Moreover, we have also checked that the ZAHB  
magnitudes given by F99 are consistent with the additional measurements 
of the TO and RGB-bump visual magnitudes. 
 
\begin{table} 
\centering 
\caption[]{Global metallicity for our selected sample of 
Galactic GCs.\label{tab3}} 
\begin{tabular}{rccccccc} 
 
NGC & [Fe/H]$_{CG}$ & HB  & log$Z_{ev}$ & log$Z_{ev}$(Vdb) \\ 
 104  &  $-$0.70 &   $-$0.99 &   $-$2.28$\pm$0.08 &  $-$2.07$\pm$0.11\\  
 288  &  $-$1.07 &   +0.98 &     $-$2.69$\pm$0.09 &  $-$2.57$\pm$0.09\\ 
 362  &  $-$1.15 &   $-$0.87 &   $-$2.69$\pm$0.12 &  $-$2.56$\pm$0.12\\ 
 1261 &  $-$1.08 &   $-$0.71 &   $-$2.66$\pm$0.07 &  $-$2.52$\pm$0.09\\ 
 1851 &  $-$1.05 &   $-$0.36 &   $-$2.64$\pm$0.07 &  $-$2.52$\pm$0.10\\ 
 1904 &  $-$1.37 &   +0.89 &     $-$2.96$\pm$0.14 &  $-$2.94$\pm$0.15\\ 
 2808 &  $-$1.13 &   $-$0.49 &   $-$2.73$\pm$0.08 &  $-$2.63$\pm$0.10\\ 
 3201 &  $-$1.23 &   +0.08 &     $-$2.88$\pm$0.08 &  $-$2.82$\pm$0.09\\ 
 4590 &  $-$1.99 &   +0.44 &     $-$3.41$\pm$0.13 &  $-$3.31$\pm$0.15\\ 
 4833 &  $-$1.58 &   +0.93 &     $-$3.17$\pm$0.15 &  $-$3.15$\pm$0.17\\ 
 5272 &  $-$1.34 &   +0.08 &     $-$2.88$\pm$0.07 &  $-$2.82$\pm$0.08\\ 
 5286 &  $-$1.49 &   +0.80 &     $-$3.06$\pm$0.17 &  $-$3.05$\pm$0.18\\ 
 5694 &  $-$1.73 &   +1.00 &     $-$3.41$\pm$0.20 &  $-$3.37$\pm$0.22\\ 
 5897 &  $-$1.59 &   +0.86 &     $-$3.12$\pm$0.22 &  $-$3.05$\pm$0.18\\ 
 5904 &  $-$1.11 &   +0.31 &     $-$2.75$\pm$0.07 &  $-$2.65$\pm$0.09\\ 
 6093 &  $-$1.44 &   +0.93 &     $-$2.94$\pm$0.12 &  $-$2.94$\pm$0.14\\ 
 6121 &  $-$1.19 &   $-$0.06 &   $-$2.71$\pm$0.12 &  $-$2.59$\pm$0.12\\ 
 6171 &  $-$0.95 &   $-$0.73 &   $-$2.54$\pm$0.09 &  $-$2.42$\pm$0.10\\ 
 6205 &  $-$1.39 &   +0.97 &     $-$3.02$\pm$0.22 &  $-$2.98$\pm$0.13\\ 
 6218 &  $-$1.26 &   +0.97 &     $-$2.86$\pm$0.12 &  $-$2.82$\pm$0.13\\ 
 6229 &  $-$1.30 &   +0.24 &     $-$2.73$\pm$0.10 &  $-$2.66$\pm$0.13\\ 
 6254 &  $-$1.41 &   +0.98 &     $-$2.98$\pm$0.15 &  $-$2.98$\pm$0.12\\ 
 6341 &  $-$2.16 &   +0.91 &     $-$3.46$\pm$0.16 &  $-$3.36$\pm$0.16\\ 
 6362 &  $-$0.96 &   $-$0.58 &   $-$2.48$\pm$0.06 &  $-$2.33$\pm$0.09\\ 
 6397 &  $-$1.82 &   +0.98 &     $-$3.07$\pm$0.12 &  $-$3.03$\pm$0.13\\ 
 6584 &  $-$1.30 &   $-$0.15 &   $-$2.84$\pm$0.12 &  $-$2.80$\pm$0.15\\ 
 6656 &  $-$1.41 &   +0.92 &     $-$3.14$\pm$0.23 &  $-$3.10$\pm$0.18\\ 
 6681 &  $-$1.31 &   +0.96 &     $-$2.84$\pm$0.09 &  $-$2.80$\pm$0.12\\ 
 6717 &  $-$1.09 &   +0.98 &     $-$2.68$\pm$0.15 &  $-$2.61$\pm$0.18\\ 
 6752 &  $-$1.42 &   +1.00 &     $-$2.92$\pm$0.11 &  $-$2.89$\pm$0.13\\ 
 6809 &  $-$1.58 &   +0.87 &     $-$3.16$\pm$0.18 &  $-$3.16$\pm$0.19\\ 
 6838 &  $-$0.70 &   $-$1.00 &   $-$2.36$\pm$0.17 &  $-$2.16$\pm$0.18\\ 
 6934 &  $-$1.30 &   +0.25 &     $-$2.89$\pm$0.13 &  $-$2.90$\pm$0.18\\ 
 7006 &  $-$1.35 &   $-$0.28 &   $-$2.98$\pm$0.22 &  $-$2.97$\pm$0.19\\ 
 7078 &  $-$2.12 &   +0.67 &     $-$3.55$\pm$0.14 &  $-$3.46$\pm$0.14\\ 
 7492 &  $-$1.38 &   +0.81 &     $-$2.90$\pm$0.13 &  $-$2.89$\pm$0.17\\ 
 6093 &  $-$1.44 &   +0.93 &     $-$3.12$\pm$0.22 &  $-$2.97$\pm$0.15\\ 
 6254 &  $-$1.41 &   +0.98 &     $-$3.27$\pm$0.15 &  $-$3.23$\pm$0.16\\ 
\end{tabular} 
\end{table} 
 
As a first step, we determine the values of
$A$=$\Delta V$(TO-BUMP)-0.566$\Delta V$(TO-ZAHB)  
from the data in Table 2, then  
we derive the global metallicity  using Eq. (4) and Eq. (5).  
The results for $Z_{ev}$ 
are listed in column (4) of Table 3, 
together with the horizontal branch morphological type  
(HB-type) \footnote{The HB-type is given by the ratio (B-R)/(B+V+R), where  
B and R denote the number of HB stars blueward and redward of the RR Lyrae  
instability strip, respectively, and V is the number of RR Lyrae stars.} from  
Harris (1996).  For the two clusters NGC 6093 and 6254, the last two lines in  
Table 3 refer to the HB level given by R99.  
 
The correlation between log$Z_{ev}$ and the  
measured [Fe/H] parameter\footnote{Note that the measured [Fe/H] values are  
taken with an uncertainty of $\pm$0.10 dex.}  
is presented in the upper panel of Fig. 7, where the solid  
line shows the {\it canonical}  
correlation log$Z$=[Fe/H]$-$1.7 for scaled-solar chemical compositions.   
One obtains that almost all the clusters present a mild overabundance of the  
global metallicity  with respect to the measured iron-to-hydrogen content.  
In terms of  $\alpha-$element enhancement [see Eq. (1)],  
we show in the bottom panel in Fig. 7 that   
the clusters with [Fe/H]$\ge-$1.7 show $f\sim$ 1.5,  
whereas the four most metal-poor clusters show $f\sim$ 3  
(dashed lines). 
 
Figure 8 displays the results based on Vdb computations.  
These models suggest slightly larger  $Z_{ev}$ values (see data listed in column (5) of  
Table 3) in comparison with our models,  
with an average difference of 0.08$\pm$0.06 dex which is  
within the uncertainty of the derived global metallicity.   
Also the $\alpha-$element enhancement  
is slightly increased, clusters  
with [Fe/H]$\ge-$1.7 now showing $f\sim$ 1.9, 
whereas the four most metal-poor clusters show $f\sim$ 3.7  
(dashed lines).  
   
We remark here that the global metallicity given in Table 3 are derived  
from the differential parameters $\Delta V$(TO-ZAHB) and  
$\Delta V$(TO-BUMP) and are  
fully independent of the measured iron-to-hydrogen content.  
On the contrary, the {\it f} enhancement factors plotted  
in the lower panel of Figs. 7 and 8 obviously depend on the adopted  
empirical [Fe/H] scale. When we consider the Zinn \& West (1984) scale  
the results are quite different. As shown in  
Fig. 9 and Fig. 10, the $\alpha-$enhancement factor is $f\sim$ 3  
(our models) or $\sim$ 3.5 (Vdb models)   
with [Fe/H]$\le -$1.3, declining toward the  
solar value $f\sim$1 at larger metallicities.  
 
The abundances of $\alpha-$elements in Population II stars   
provide fundamental constraints to the chemical evolution  
models of the Galaxy.  
Several spectroscopic studies have been devoted to  
this problem (see Carretta, Gratton \& Sneden 2000 for discussion and  
references).  
As a whole, $\alpha-$elements appear to be  
overabundant by a constant factor $\sim$2-3  
in metal-poor stars, declining toward scaled solar ratios
around [Fe/H]=$-$1.0. As the run of  
[$\alpha$/Fe]\footnote{[$\alpha$/Fe]=log$f$.}  
with [Fe/H] is required to study the chronology of the Galactic halo  
formation (Matteucci \& Francois 1992), we notice 
that mild variations of the empirical [Fe/H] scale may lead to quite  
different pictures for the dependence of [$\alpha$/Fe] on [Fe/H], and
different [Fe/H] values for 
the transition toward scaled-solar ratios.   
 
Both the distribution and absolute values of the  
$\alpha-$element enhancement suggested by our method together with   
the Zinn \& West (1984) [Fe/H] scale 
appear in fine agreement with the empirical results by Carney (1996). 
On the contrary, the results obtained using 
the Carretta \& Gratton (1997) [Fe/H] scale do not appear
fully supported by current spectroscopic measurements.  
Even though to investigate the accuracy of current  
empirical [Fe/H] scales is 
out of the aim of the present paper, we wish to emphasize that this is still a controversial 
issue. We also wish to notice that recently Vandenberg (2000) and Bergbusch \& 
Vandenberg (2001) have provided some additional evidence, based on the comparison between their 
theoretical isochrones and CMDs of Galactic GCs, that the metallicities determined by Carretta \& 
Gratton (1997), mainly for the intermediate-metallicity clusters, are too high, values 
close to the Zinn \& West (1984) scale being more favourite. 
Present analysis provides independent support to their claim. 
 
Before closing this section, it is worthwhile to emphasize  
that our method, being based only on well defined observational features  
of the CMD, appears  quite suitable for deriving accurate estimates of the GC global 
metallicity. It uses only differential quantities,  i.e. 
visual magnitude differences, and its results are therefore  
insensitive to uncertainties on the 
zero point of the CMD photometric calibration as well as on the cluster  
distance and reddening. 
 
It is also worth noticing that the method relies on observables whose
properties 
are not significantly affected  
by non-canonical processes occurring along the RGB, like deep-mixing processes,  
which are currently invoked 
to explain the chemical abundance patterns observed in RGB stars of many Galactic GCs 
(Sweigart \& Mengel 1979, Langer, Hoffman \& Sneden 1993). As suggested by  
theoretical predictions (Mestel 1957) and spectroscopic data  
(Charbonnel, Brown, \& Wallerstein 1998, and references therein),  
molecular weight gradients, such as the H-discontinuity left over in the
envelope of 
RGB stars by the first dredge up, should be able to  
inhibit these non-canonical mixing processes, which are therefore 
expected to be efficient only after the RGB bump.

\section{Summary}

The main results obtained in present work can be summarized as follows: 
 
\begin{itemize} 
 
\item	We have developed an homogeneous theoretical scenario, based on updated stellar 
models including atomic diffusion, computed 
for a wide metallicity range; this allows us 
to predict the behaviour with the cluster metallicity and age of relevant  
observational features of the CM diagram, such as the Turn Off magnitude and the  
brightness of both the ZAHB at the level of the RR Lyrae instability strip,  
and the RGB-bump; 
 
\item	we define a new observable based on the visual magnitude difference between the 
TO and the ZAHB $\Delta M_V$(TO-ZAHB), and the TO and the RGB-bump $\Delta M_V$(TO-BUMP),  
namely, $A= \Delta M_V$(TO-BUMP)$-$0.566$\Delta M_V$(TO-ZAHB).
We show that the parameter $A$ does not depend at all on the cluster age, 
whereas it does strongly correlate with the {\it global} cluster metallicity.  
The calibration 
of the parameter $A$ as a function of $Z$ is provided; 
 
\item	by using this calibration, together with a large observational database of  
Galactic GCs, we estimate the global metallicity of a large sample of GCs. 
Our estimates are independent of empirical metallicity 
measurements, but strongly rely on the accuracy and reliability of the adopted theoretical 
models. By using our global metallicity determinations, together with  
presently available [Fe/H] determinations
(Zinn \& West 1984 or Carretta \& Gratton 1997), and using the relation 
provided by Salaris et al. (1993) that connects global metallicity to  
[Fe/H] and $\alpha-$element 
enhancement, we are able to provide an estimate of the [$\alpha$/Fe]  
ratio for the individual clusters in our sample; 
 
\item	the trend of [$\alpha$/Fe] with [Fe/H] 
we predict when using the Zinn \& West metallicity scale, appears  
in fine agreement with current 
spectroscopical measurements, which suggest a rather constant  
overabundance of $\alpha-$elements in metal-poor stars, and a decline 
toward the solar ratios around [Fe/H]$\approx -$1.0. This picture  
is not supported when we use the  
Carretta \& Gratton [Fe/H] scale; 
 
\item	the calibration of the $A-Z_{ev}$ relation is model dependent,  
in the sense that the Vdb models, which are based on  
different input physics, yields a slightly different formulation of the  
parameter $A$, given by $A$=$\Delta M_V$(TO-BUMP)$-$0.487$\Delta M_V$(TO-ZAHB).  
However, the resulting $Z_{ev}$ values are only 0.08$\pm$0.06 dex larger than the  
results based on our computations; 
 
\item the good agreement between the global metallicity estimates provided  
by our models accounting for atomic diffusion, and the ones obtained 
by employing Vdb models which neglect atomic diffusion, shows 
that the theoretical calibration of the $A$ parameter is only marginally 
affected by the efficiency of this non-canonical process; 
 
\item the present analysis clearly demonstrates that the our newly 
defined observational parameter, independent of  
the uncertainty on the zero point of the photometric calibration as well as on the cluster  
distance and reddening, is a powerful indicator of globular cluster  
global metal abundances.  
  
\end{itemize}

{\bf Acknowledgments:}  
 
We warmly thank D. Vandenberg, P. Bergbusch  and   
M. Salaris for sending us data about their theoretical models.  
We gratefully also acknowledge M. Salaris for an accurate reading of an early
version of this manuscript and for his helpful comments.
We thank the anonymous referee for several helpful suggestions and comments.   
One of us (C.S.) wishes to acknowledge the 
Max Planck Institut f\"ur Astrophysik for the kind hospitality during  
the early phase of this work. Financial support for this work was provided by  
MIUR-COFIN2000, under the scientific project "Stellar observables of  
cosmological relevance".

\end{document}